\documentclass[a4paper,12pt]{article}

\usepackage[T2A]{fontenc}
\usepackage[english]{babel}
\usepackage{amsmath}
\usepackage{amssymb}
\usepackage{cite}
\usepackage[unicode,linktocpage=true,plainpages=false,pdfpagelabels=false]{hyperref}

\newcommand{\dd}{\partial}
\newcommand{\m}{\mu}
\newcommand{\n}{\nu}
\newcommand{\ls}{\left(}
\newcommand{\rs}{\right)}
\newcommand{\ka}{\varkappa}
\newcommand{\ff}{\varphi}
\newcommand{\ta}{\tau}
\newcommand{\al}{\alpha}
\newcommand{\te}{\theta}

\newcommand{\disn}[2]{$$\displaylines{\refstepcounter{equation}%
            \label{#1}\hskip 1em minus 1em #2\hfilneg}$$}
\newcommand{\nom}{\hfil\hskip 1em minus 1em (\theequation)}

\begin{document}

\title{Gravity as embedding theory\\ and the distribution of matter in galaxies}

\author{
S.~A.~Paston\thanks{E-mail: pastonsergey@gmail.com},
A.~D.~Kapustin\thanks{E-mail: sashakapusta96@gmail.com}\\
{\it Saint Petersburg State University, Saint Petersburg, Russia}
}
\date{\vskip 15mm}
\maketitle

\begin{abstract}
The description of gravity in the form of an embedding theory is based on the hypothesis that our space-time is a four-dimensional surface in a flat ten-dimensional space. The choice of standard Einstein-Hilbert action leads in this case to more general field equations than Einstein's equations. By writing them in the form of Einstein's equations with the contribution of additional fictitious matter, one can try to interpret this matter as dark matter. In order to study the behavior of this fictitious matter near the centers of real galaxies, we discuss an analytical method of obtaining corresponding matter density profiles. This method is based on the consideration of the distribution function of particles over all possible trajectories and allows us to estimate the type (core or cusp) of the emerging density profile.
\end{abstract}

\newpage

\section{Introduction}
In the usual description of gravity in the framework of general relativity, four-dimensional space-time is a pseudo-Riemannian space. As an example of a Riemannian or pseudo-Riemannian space, one can consider a $d$-dimensional surface in a flat space of a larger number of dimensions, assuming that the metric of this surface is induced; i.e., distances between close points on a surface are defined as the distance between these points in the surrounding space. In particular, an easily representable example of a Riemannian space is a two-dimensional surface in three-dimensional space.

It turns out that a surface in a flat ambient space is not only a special case of a Riemannian space, but, to a certain extent, the consideration of an arbitrary Riemannian space can be replaced by the consideration of such a surface. This conclusion allows us to make the Janet-Cartan-Friedman theorem \cite{fridman61}, which says, in particular, that an arbitrary pseudo-Riemannian space of dimensiond $d$ can be locally isometrically embedded in a flat pseudo-Euclidean ambient space
with a suitable signature and dimension $d(d+1)/2$.
This means that, for a four-dimensional space-time, it is sufficient to take a ten-dimensional space as the enclosing one. In this case, it suffices to assume the existence of a single timelike direction in it, i.e., to consider that enclosing pseudo-Euclidean space $R^{1,9}$.

Since the consideration of space-time as a pseudo-Riemannian space can, with some reservations, be replaced by the consideration of a four-dimensional
surface in a ten-dimensional space $R^{1,9}$, the following idea arises: when describing gravity as independent variables, use not the metric field $g_{\m\n}(x)$, but the variables describing the surface. As such variables, it is most convenient to choose the embedding function $y^a(x^\m)$
(where $a,b,\ldots=0,\ldots,9$) that displays
 \disn{v3.2}{
y:R^4\longrightarrow R^{1,9}.
\nom}
As a result, gravity is described as the dynamics of a three-dimensional space, considered similarly to the dynamics of a point particle, which corresponds to a world line, and to the dynamics of a string, to which a two-dimensional surface corresponds in Minkowski space. This analogy is one of the main motivations for the description of gravity under discussion (often called the embedding theory), which was first proposed in \cite{regge}. In Section~2, we briefly describe this theory of gravity and show that its equations of motion can be written as \cite{pavsic85} in the form of a set of Einstein's equations with the contribution of additional fictitious matter, as well as equations that somehow limit the properties of this matter.

This possibility allows one to try to interpret the fictitious matter that arises within the framework of the embedding theory as dark matter, whose existence hypothesis explains the observations in accordance with the $\Lambda$CDM model; see, for example, \cite{gorbrub1}. Within its framework, dark matter can be considered a non-relativistic dustlike matter, which generates the same
gravitational field as ordinary matter, and the non-gravitational interaction of dark matter with ordinary matter is either absent or turns out to be undetectable weak. Because numerous ongoing attempts to directly detect dark matter have not yet yielded results \cite{1509.08767,1604.00014}, there is every reason to consider dark matter as a purely gravitational effect associated with some modification of the theory of gravity.

In addition to the embedding theory, other variants of modifications containing non-Einstein solutions can be used for this purpose. Particularly well suited for this are the modifications arising, like the embedding theory, as a result of a change of the independent variable containing differentiation in the action of general relativity (see details in \cite{statja60}). Among such theories, mimetic gravity is the most widely known \cite{mukhanov,Golovnev201439} (see also overview \cite{mimetic-review17}). In order to successfully explain the effects associated with dark matter on different scales, it is necessary that the fictitious matter that arises within the framework of a particular modification of gravity has a sufficiently large number of degrees of freedom. For mimetic gravity, this amount is insufficient; therefore, in order to successfully explain observations, the theory has to be complicated by introducing additional terms \cite{mimetic-review17}. For the embedding theory, the number of degrees of freedom of fictitious matter is large enough, which makes it possible to discuss the effects both on cosmological scales \cite{davids01,statja26} and on the scale of galaxies. However, a comparison with observations in the latter case requires additional studies of the behavior of fictitious matter in the nonrelativistic limit, in which it behaves like dusty matter with some self-action \cite{statja68}.

One question that needs to be studied is what profile forms the distribution of dark matter in galaxies if dark matter is the fictitious matter of the embedding theory. This question is of particular interest in connection with the existence of well-known \emph{core-cusp} problem. This problem lies in the discrepancy between the observed data on the distribution of dark matter near the centers of galaxies, in which the density of matter is sufficiently smooth and forms the \emph{core}, with the results of numerical simulations of the process of formation of cosmic structures, in which the initial distribution of matter comes in the process of evolution to a singular distribution of matter, called the \emph{cusp}. The classical result of numerical simulation work is the universal density profile \cite{Navarro_1996}, often called the NFW distribution, in which the density of the number of particles in the vicinity of the center of the galaxy is
singular and behaves like $r^{\alpha}$ with $\alpha = -1$. This profile is consistently reproduced in simulations corresponding to the formation of structures from noninteracting dark matter; see, for example, \cite{galaxies10010005}. The solution to the core-cusp problem can be sought by introducing some dark matter self-action (see, for example, the work \cite{1705.02358}), since the self-action can prevent the excessive thickening of dark matter in the center of galaxies.
Therefore, the description of dark matter as a fictitious matter of the embedding theory, which has some interaction in the nonrelativistic limit, seems quite promising.

It is necessary to learn how to evaluate how the appearance of self-action affects the emerging profile of the distribution of matter in the galaxy. In order to avoid conducting volumetric numerical simulations when receiving an answer to this question, the task arises to achieve at the analytical level an understanding of when which distribution -- core or cusp -- occurs. First and foremost, this question must be answered in the absence of self-action and then, when studying the properties of fictitious matter with a specific type of self-action, one can add the consideration of this self-action as some modification of the problem. In Section~3, we describe the simple analytical method proposed in \cite{statja73} for determining the asymptotics at zero for the radial dependence of the density of dusty matter in a galaxy by considering the distribution function of particles along possible trajectories. Basically, the discussion is limited to the average static and spherically symmetric distribution; however, the movement of each specific particle breaks the spherical symmetry.

\section{Gravity as an embedding theory}\label{razdvlog}
When describing gravity in the form of an embedding theory, the independent variable describing the gravitational field is the embedding function \eqref{v3.2}, and the metric is expressed in terms of it by the induced metric formula
\disn{spn1}{
g_{\m\n}=(\dd_\m y^a)(\dd_\n y^b)\eta_{ab},
\nom}
where $\eta_{ab}$ -- is a flat metric of the ambient space $R^{1,9}$.
As an action of the theory, the Einstein-Hilbert action, the standard for general relativity, is taken
 \disn{v3.3}{
S=-\frac{1}{2\ka}\int d^4x\, \sqrt{-g}\;R,
\nom}
into which the metric is substituted \eqref{spn1}.
As a result of varying this action on $y^a(x)$ the Regge-Teitelboim equations arise~\cite{regge}
\disn{15}{
D_\m\Bigl(
\ls G^{\m\n}-\ka\, T^{\m\n} \rs \dd_\n y^a\Bigr)=0.
\nom}
These equations are more general than Einstein's
$G^{\m\n}=\ka\, T^{\m\n}$; i.e., any solution of the Einstein equations will be a solution to the equations of the embedding theory, but not vice versa. Thus, an embedding theory falling into the class of theories obtained from general relativity by a differential change of variable \cite{statja60} contains non-Einstein solutions. After article \cite{regge}, the ideas of embedding theory were critically discussed in the work \cite{deser}, and subsequently they were used quite a lot to describe gravity, including in connection with its quantization; see, for example, \cite{tapia,maia89,frtap,davkar,statja25,faddeev,statja33}.
A detailed list of literature related to embedding theory and related issues can be found in the review \cite{tapiaob}.

Equations of motion \eqref{15} embedding theories can be rewritten \cite{pavsic85} as a set of equations:
 \disn{s4}{
G^{\m\n}=\ka \ls T^{\m\n}+\ta^{\m\n}\rs,\qquad
D_\m\Bigl(\ta^{\m\n}\dd_\n y^a\Bigr)=0.
\nom}
The first is the Einstein equation containing the additional contribution of the energy-momentum tensor $\ta^{\m\n}$
of some fictitious matter. The second can be perceived as an equation that limits the possible behavior of quantity $\ta^{\m\n}$
and, hence, as an equation for the motion of this fictitious matter. When writing the equations of motion of the embedding theory in the form \eqref{s4}, the analysis of the properties of fictitious matter is simplified \cite{statja68}.

\section{Matter density profile analysis}\label{razdcorecusp}
When modeling the distribution of matter in a galaxy, in the first approximation, deviations from spherical symmetry can be neglected. Let us consider, on average, a spherically symmetric and time-independent distribution of dusty matter consisting of massive particles interacting only gravitationally and moving along limited trajectories. We assume that the particles move with nonrelativistic velocities and the density of matter $\rho(x)$ is small enough. Then it is related to the
static spherically symmetric gravitational potential $\varphi(x)$ Poisson equation
\begin{equation}
\label{poisson}
	\Delta \varphi(x) = 4 \pi G \rho(x),
\end{equation}
where $G$ is Newton's gravitational constant.

When moving in a spherically symmetric potential for each particle, its total energy E  and angular momentum $L_k$ ($i,k,\ldots=1,2,3$) are conserved. In this case, the movement of each particle occurs along a flat orbit, the plane of which passes through the center of symmetry. When considering only finite motion, the change in the radial coordinate will be periodic, but the orbit may or may not be closed. Each orbit can be uniquely specified using the specific (per unit mass $m$) energy $\varepsilon = {E}/{m}$;
specific angular momentum $\ell_k = {L_k}/{m}$; and directions $\tau_k$, which determines the
position of the orbit in the plane perpendicular to the angular momentum vector $L_k$. For a complete description of the motion of a particle, one must also introduce a scalar parameter $\gamma$, which specifies at the initial moment of time the phase of the particle during
periodic orbital motion. As a result, the motion of a particular particle is described by the function $\hat{x}_m \left( t, \varepsilon, \ell_k, \tau_l, \gamma \right)$, where $t$ is time.

By introducing the particle distribution function $f$, one can use it to write the expression for the density in the form
\begin{equation}
\label{rho}
	\rho(x_m) = m \int d \varepsilon \,d^3 \ell \,d \tau \,d \gamma \ f \left(  \varepsilon, \ell_k, \tau_l, \gamma  \right) \delta \left( x_m -  \hat{x}_m \left( t, \varepsilon, \ell_k, \tau_l, \gamma \right) \right).
\end{equation}
Given the assumed spherical symmetry and time independence of the density of matter $\rho$, it can be written in the form (see details in \cite{statja73})
\begin{equation}\label{rho5a}
\rho(r) = \frac{m}{2\pi r^2} \int d \varepsilon \, \int\limits_0^\infty d\ell\,
\frac{\hat f(\varepsilon, \ell)\,\te \left( 2\varepsilon - 2\varphi(r) - \frac{\ell^2}{r^2}\right)}{\hat{T} \left( \varepsilon, \ell \right)\sqrt{2\varepsilon - 2\varphi(r) - \frac{\ell^2}{r^2}}},
\end{equation}
where $\hat{T} \left( \varepsilon, \ell \right)$ is the period of radial motion of the particle along the orbit,
$r=\sqrt{x_k x_k}$, $\ell=\sqrt{\ell_k\ell_k}$, $\te(x)$ is the Heaviside theta function and the notation
\begin{equation}\label{sp11}
\hat f(\varepsilon, \ell)=
\int \limits_{S_\ell} d^2 \ell  \,
\int d \tau \,d \gamma\, f(\varepsilon, \ell_k, \tau_l, \gamma),
\end{equation}
and $S_\ell$ denotes a sphere of radius $\ell$.
Value $\hat f(\varepsilon, \ell)$ the meaning of the distribution function of particles in terms of specific energy and modulus of angular momentum.

Consider the behavior of the density \eqref{rho5a} in asymptotics $r \to 0$. Due to the fact that the gravitational potential is related to the nonnegative density of matter by Eq.~\eqref{poisson}, value $-\varphi(r)$ cannot at $r \to 0$ grow faster or the same as ${1}/{r^2}$.
Therefore, for small $r$, term ${\ell^2}/{r^2}$ turns out to be dominant in the argument of $\theta$-functions, and hence the contribution to the integral over $\ell$ gives only the range of small values
\begin{equation}
\ell\le r\sqrt{2\varepsilon - 2\varphi(r)}.
\end{equation}
As a result, the asymptotics $\rho(r)$ at $r\to0$
is determined by the asymptotics at $\ell\to0$ included in \eqref{rho5a} functions $\hat f(\varepsilon, \ell)$ and $\hat{T}(\varepsilon,\ell)$.
The period of radial motion of a particle in orbit $\hat{T}(\varepsilon,\ell)$ has at $\ell\to0$ final limit $\hat{T}(\varepsilon,0)$, with the possible exception of giving a small contribution to very small values of specific energy $\varepsilon$, which corresponds to particles that are almost at rest in the center. Nothing definite can be said in advance about the distribution function $\hat f(\varepsilon, \ell)$, so we consider various possible variants of its asymptotics.

Let us first assume that $\hat f(\varepsilon, \ell)$ has at $\ell\to0$ the final limit for which at least in some area of change $\varepsilon$
will be $\hat f(\varepsilon, 0)\ne0$. In this case, when integrating
into \eqref{rho5a} on $\ell$, make a change of variables $\ell=r\tilde\ell$, which leads to the asymptotics
\begin{equation}\label{rho5}
\rho(r) = \frac{m}{2\pi r} \int d \varepsilon \, d\tilde\ell\,
\frac{\hat f(\varepsilon, 0)\,\theta \left( 2\varepsilon - 2\varphi(r) - \tilde\ell^2\right)}{\hat{T} \left( \varepsilon, 0 \right)
\sqrt{2\varepsilon - 2\varphi(r) - \tilde\ell^2}}=
\frac{m}{4 r} \int d \varepsilon \,\frac{\hat f(\varepsilon, 0)}{\hat{T} \left( \varepsilon, 0 \right)}.
\end{equation}
As can be seen, the density of the number of particles at $r\to0$ in this case turns out to be singular, and it behaves as $1/r$
(note that $\hat f(\varepsilon, 0)>0$, so that the integral in \eqref{rho5} does not vanish). This exactly corresponds to the profile of the type \emph{cusp} with $\al=-1$.
Note that the gravitational potential in this case remains finite at $r=0$, but has a break at this point; i.e., $\varphi'(0)\ne0$.

Now suppose that the distribution function $\hat f(\varepsilon, \ell)$ at $\ell\to0$ can be expanded in series $\ell$, but its zero expansion term turns out to be equal to zero for all $\varepsilon$; i.e., at $\ell\to0$,
\begin{equation}\label{sp4}
\hat f(\varepsilon, \ell)\approx \hat f'(\varepsilon, 0)\ell,
\end{equation}
where the prime means the derivative with respect to $\ell$. In this case, arguing similarly, instead of \eqref{rho5}, we get
\begin{equation}\label{rho6}
\rho(r) = \frac{m}{2\pi} \int d \varepsilon \, d\tilde\ell\,
\frac{\hat f'(\varepsilon, 0)\tilde \ell\,\theta \left( 2\varepsilon - 2\varphi(r) - \tilde\ell^2\right)}{\hat{T} \left( \varepsilon, 0 \right)
\sqrt{2\varepsilon - 2\varphi(r) - \tilde\ell^2}}=
\frac{m}{2\pi} \int d \varepsilon \,\frac{\hat f'(\varepsilon, 0)}{\hat{T} \left( \varepsilon, 0 \right)}\sqrt{2\varepsilon - 2\varphi(r)}.
\end{equation}
Assuming limited value $\varphi(0)$ of the gravitational potential at zero (which takes place even in the above \emph{cusp} case), we conclude that, when considering the asymptotics $r\to0$ in the last expression in \eqref{rho6}, $\varphi(r)$ can be replaced with $\varphi(0)$.
This means that in this case the dependence of the density on the radius corresponds to a profile of the type \emph{core}.

Thus, a fairly simple analytical reasoning allows us to conclude that the matter density profile formed by matter in the center of the galaxy is determined by the presence or absence of a zero term in the expansion in
terms of $\ell$,  functions $\hat f(\varepsilon, \ell)$  at $\ell=0$. However, given
the positive $\hat f(\varepsilon, \ell)$, it suffices to analyze the presence
or absence of the zero expansion term for $\ell=0$  for the distribution function
\disn{sp15}{
\hat f(\ell)=\int d\varepsilon \hat f(\varepsilon, \ell)
\nom}
of particles only according to modulo specific angular momentum. It can be shown that, at the initial moment of time, when the distribution of particles in coordinates and velocities has a certain random form, this distribution function for $\ell\to0$ has behavior
\disn{sp11a}{
\hat f(\ell)\approx C\ell,
\nom}
see details in \cite{statja73}.

If we consider the situation whena  cloud of particles forms a stationary configuration on average in a predetermined spherically symmetric gravitational
potential $\ff(x_i)$, then for each particle in its motion, the
angular momentum is conserved. Then the defined \eqref{sp15} distribution function $\hat f(\ell)$ does not change over
time, and its behavior \eqref{sp11a} will also take place for the formed stationary configuration of particles. Since such asymptotics corresponds to \eqref{sp4}, we can conclude that, in the case under consideration, a matter density profile of the type \emph{core} arises. Such a situation, when the gravitational potential remains sufficiently spherically symmetric all the time during the formation of the density profile of the matter under consideration, can be realized when the gravitational potential is mainly created by dark matter. At the same time, dark matter can obey its own laws (for example, having some specific self-action), determining whether it forms a \emph{cusp} or \emph{core}, while we study the formation of the density profile of ordinary matter and obtain for it a distribution of the \emph{core} type.

An alternative situation is the case when the cloud of particles forms a stationary configuration on average in the absence of a given spherically symmetric gravitational potential. Then the gravitational potential is formed simultaneously with the emergence of a stationary configuration and, in the process of such a formation, noticeable deviations from spherical symmetry can occur. This situation can be realized if we study the formation of a profile for dark matter particles, assuming that it is an ordinary dusty matter without any self-action. This formulation of the problem is close to that used in the abovementioned numerical simulations.

In this case, the particle trajectories will no longer correspond exactly to the trajectories described above, characterized by the parameters $\varepsilon, \ell_k, \tau_l, \gamma$. However, if
the spherical symmetry is not broken too much, the deviations from such trajectories will not be very large and the particles can still be described
by the parameters $\varepsilon, \ell_k, \tau_l, \gamma$, but already considering that not all of
them are preserved over time. In particular, for each particle, the angular momentum defined relative to the center of the emerging stationary configuration is no longer preserved, since it can change under the action of forces acting from the centers of small concentrations of particles, which can eventually form satellites. As a result, in contrast to the case described above, the distribution function $f(\varepsilon, \ell_k, \tau_l, \gamma)$, which means $\hat f(\ell)$, may change over time. As a consequence,
its asymptotic behavior may change with time as $\ell=0$, which determines whether a \emph{core} or \emph{cusp} is formed.
As a result of the flow of particles in the space of the modulus of the specific angular momentum to
the point $\ell=0$, distribution function over time $\hat f(\ell)$ can change asymptotics \eqref{sp11a} on
\eqref{sp11a} on
\disn{sp17}{
\hat f(\ell)\approx \hat f(0)+C\ell
\nom}
with $\hat f(0)>0$.
Since, as was mentioned above, this corresponds to the matter profile of the \emph{cusp} type, we find that such a profile can arise during the formation of a stationary configuration of particles in the presence of deviations from spherical symmetry.

{\bf Acknowledgements.}
This work was supported by the Russian Foundation for Basic Research, grant no.~20-01-00081.


\end{document}